# Propagation of Light in a Hot and Dense Medium

Samina S. Masood[1]

Department of Physics, University of Houston Clear Lake,
Houston TX 77058

Photons, as quanta of electromagnetic fields, determine the electromagnetic properties of an extremely hot and dense medium. Considering the properties of a photon in the interacting medium of charged particles, we explicitly calculate the electromagnetic properties such as the electric permittivity, magnetic permeability, refractive index and the propagation speed of electromagnetic signals in extremely hot and dense background in cosmos. Photons acquire dynamically generated mass in a medium. The screening mass of photon, Debye shielding length and the plasma frequency are functions of statistical parameters of the medium and are studied in detail. We study the properties of the propagating particles in astrophysical systems of distinct statistical conditions. The modifications in the medium properties lead to the equation of state of the system. We mainly calculate all these parameters for the extremely high temperature conditions of the early universe.

## 1. INTRODUCTION

We re-investigate the behavior of the first generation of leptons and corresponding electromagnetic field in extremely hot and dense medium of electrons below the electroweak temperature. This system is composed of different types of particles but the significant contribution comes from the light particles which have masses lower than the temperature in the early universe. Overall behavior of particles in a medium is a net result of interaction of propagating particles with the medium. The background corrections become much more significant at high temperatures and densities where the propagating particles can modify the properties of the medium [1-5]. We consider the light mass particles in a heat bath of electrons and photons at temperature below the decoupling temperature. Thus the higher generations of leptons are not expected to affect or be affected significantly by temperature which are much below their masses. Radiative background corrections due to the heavy intermediate vector bosons of electroweak interactions are also suppressed because of their heavy mass. Therefore, we study the background contribution of radiation while it is interacting with matter and temperature of hot electrons is below 2 MeV, the decoupling temperature. The system is considered to be in thermal equilibrium in specified regions of the stellar bodies.

In this paper, we study a pure gas of electrons and photons that can be converted in to electromagnetic plasma of photons and electrons at high temperatures which are sufficiently smaller than the W and Z masses. Therefore, the background contribution is coming from the lepton and photon propagators and the neutrinos are not yet decoupled because electrons have lower than neutrino decoupling energy. However, electromagnetic properties of the propagating neutrinos ( if they have mass) will be modified at high temperatures due to the electron induced magnetic moment. The heatbath with a high concentration of electrons and photons can still be considered as a relativistic plasma of electrons and photons (for μ << m). Photon acquires a temperature dependent screening mass and Debye screening length can be calculated from the longitudinal component of the vacuum polarization tensor. Electromagnetic properties of the medium are modified. Massive photon can then be treated as a Plasmon. The Plasmon at higher energy can decay in to a neutrino-antineutrino pair which can couple with the Plasmon through electrons as a higher order effect. The first order radiative corrections to electromagnetic vertex are shown in Figure (1). These diagrams contribute to the magnetic moment of electron (Figure 1a) and neutrino (Figure (1b and1c). Electron induces a nonzero magnetic moment to neutrino due to the interaction of electron with neutrino in $e^- \nu_e \rightarrow e^- \nu_e$ in the minimal standard models with very tiny mass of neutrino. Thermal background corrections due to the second or higher generation of particles will always be suppressed even if those particles are injected in the medium from outside.

[1] Electronic address: masood@uhcl.edu

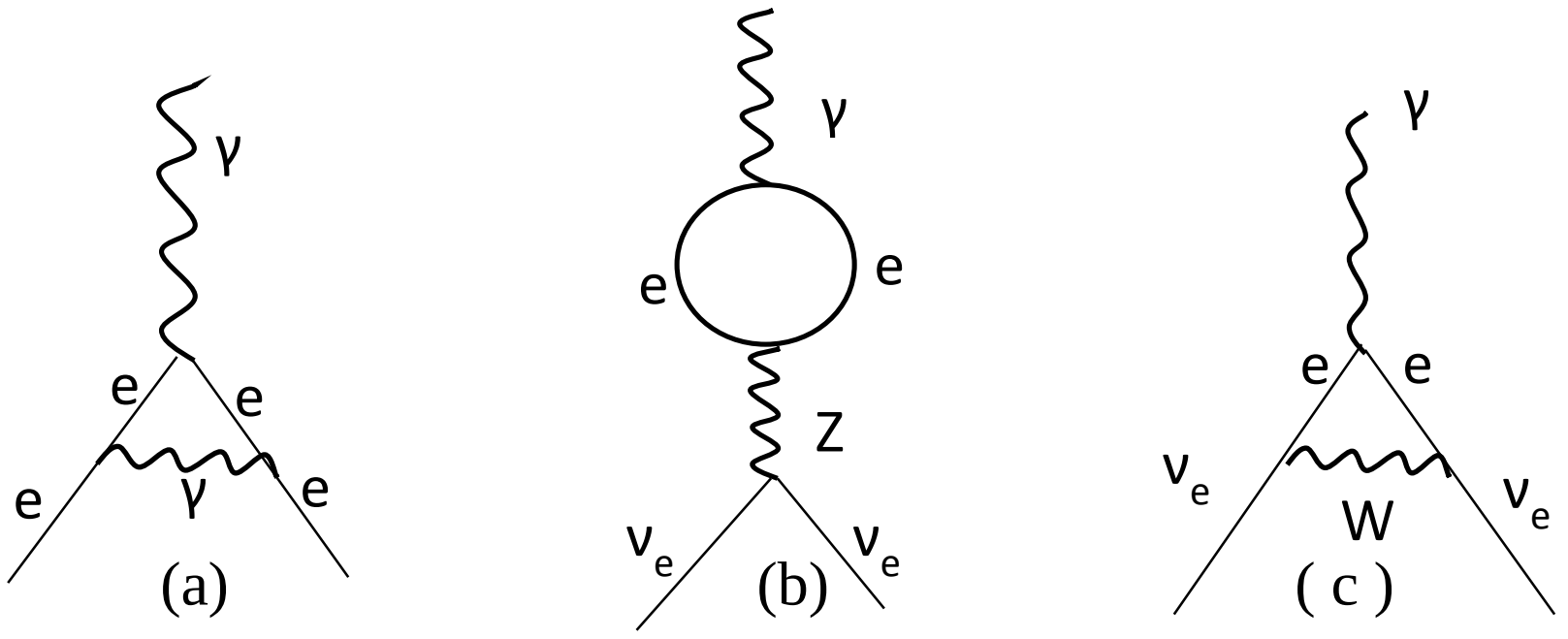

*Figure 1: First order radiative corrections to plasmons in the electroweak model results in to electrons or neutrinos or the interaction of leptons with the magnetic field. Radiative corrections to electromagnetic vertex (a), the tadpole diagram (b) correspond to neutral current and gives nonzero contribution for asymmetric combination of electrons-positron backgroud only. The bubble diagram (c) gives a major contribution to magnetic moment of neutrino.*

## 2. Calculational Scheme

The properties of electrons as electromagnetically interacting particles in an extremely hot and dense system are studied using the renormalization scheme of quantum electrodynamics (QED) in statistical media for different ranges of temperature and chemical potential [7-13]. Temperatures and chemical potential are in the range where we deal with the real particles only and the real-time formalism is used. This formalism is valid in a heat bath of real particles below the decoupling temperatures. All of the Feynman rules of QED remain unchanged. The statistical effects are included through the statistical distribution functions. We incorporate the Bose-Einstein distribution function for massless vector bosons. Fermi-Dirac distribution function is given by [6]

$$n_F(p,\mu) = \left[ \frac{\theta(p_0-\mu)}{e^{\beta(p_0-\mu)}+1} + \frac{\theta(p_0+\mu)}{e^{\beta(p_0+\mu)}+1} \right] \qquad (1)$$

We consider, in Eq. (1), a closed system with equal and opposite chemical potential to the corresponding antifermions in a CP symmetric background. First term in parenthesis corresponds to particle distribution, whereas the second term corresponds to antiparticles distribution in hot and dense medium. It is convenient to expand the distribution functions of particle and antiparticle in powers of mβ (for a constant chemical potential), where m is the mass of the corresponding particles and β=1/T. All the statistical parameters μ, T and B are expressed in units of m, the electron mass. In a heat bath of electrons, at very high temperatures, properties of electrons change corresponding to temperature and density of the system. The physically measureable values of electron mass, charge and wavefunction of electrons in a medium are calculated as renormalization constants [5] of QED in a hot and dense heat bath for different ranges of temperature and chemical potential. Without getting in to details of calculations, we use the physically measureable parameters of the propagating particle with the renormalization constants of QED to determine the electromagnetic properties of the medium as a relativistic plasma. QED renormalization constants, combining with the bare parameters give the physically measureable quantities of the system. These renormalized finite quantities correspond to the physically measureable values of the parameters such as electron mass $m^R$ [7,8], charge [9-11], wavefunction [12] and the magnetic moments [13-20]. We can then replace the rest mass of electron $m^0$ by the physically measureable renormalized mass as:

$$m^R = m^0 \left(1+\frac{\delta m}{m}\right) \qquad (2a)$$

$$m^R \equiv m^{Phys}, \qquad (2b)$$

whereas, the corresponding relations between renormalized wavefunction of electron with that of the corresponding vacuum value is given as

$$\psi^R = \frac{1}{\sqrt{Z_2}}\psi^0 \qquad (3a)$$

$$\psi^R = \psi^{Phys} \qquad (3b)$$

such that the probability of finding particles in certain states becomes a function of the statistical parameters of the medium. The electromagnetic fields is expressed as

$$A_\mu{}^R = \frac{1}{\sqrt{Z_3}} A_\mu{}^0 \qquad (4a)$$

$$A_\mu{}^R = A_\mu{}^{Phys} \qquad (4b)$$

and the physical mass Eq.(2a), wavefunction Eq.(3a) and the electromagnetic fields Eq.(4a) give the physically measureable values of the corresponding parameters.The QED Lagrangian of such a system can then be written as

$$L = -\frac{1}{4} Z_3 F_{\alpha\mu\nu} F_\alpha^{\mu\nu} + iZ_2 \overline{\psi}_R \gamma_\mu D^\mu \psi_R + Z_2 \overline{\psi_R} m_0 \psi_R + e_0 Z_2 \sqrt{Z}_3 \overline{\psi_R} \gamma_\mu A^\mu{}_R \psi_R \qquad (5)$$

In this scheme of calculations, the renormalization constants of QED are considered to be the effective parameters of the theory. The renormalization constants of QED give the physical mass and the charge of electrons and the corresponding wavefunction at finite temperature and density. The vacuum polarization tensor $\Pi_{\mu\nu}$ for such a system can be written by replacing the photon and electron propagator in vacuum by the one in the medium in real-time formalism such that:

$$\pi_{\mu\nu}(K,\mu) = ie^2 \int \frac{d^4p}{(4\pi)^4} Tr\left\{\gamma_\mu(\not{p} + \not{K} + m)\gamma_\nu(\not{p} + m)\right\}\left[\frac{1}{(p+K)^2 - m^2} + \Gamma_F(p+K,\mu)\right] \left[\frac{1}{p^2 - m^2} + \Gamma_F(p,\mu)\right] \quad (6)$$

With [6]

$$\Gamma_F(p,\mu) = 2\pi i\delta(p^2 - m^2)[\theta(p_0)n_F(p,\mu) + \theta(-p_0)n_F(p,-\mu)]\,, \quad (7)$$

whereas,

$$K^2 = \omega^2 - k^2, \qquad \omega = K_\alpha u^\alpha. \quad (8)$$

$K^2 = 0$ is in vacuum due to the transverse nature of light which assures the absence of longitudinal component as well as the masslessness of photon. $u^\alpha$is the 4-velocity of the heatbath.

The vacuum part $\pi_{\mu\nu}^{T=0}(K)$ and the medium contribution $\pi_{\mu\nu}^{\beta}(K,\mu)$ to Eq. (6) can be written as

$$\pi_{\mu\nu}(K) = \pi_{\mu\nu}^{T=0}(K)+\pi_{\mu\nu}^{\beta}(K,\mu) \quad (9)$$

with

$$\pi_{\mu\nu}^{\beta}(K,\mu) = -\frac{2\pi e^2}{2}\int \frac{d^4p}{(4\pi)^4} Tr\left\{\gamma_\mu(\not{p} + \not{K} + m)\gamma_\mu(\not{p} + m)\right\} \left[\frac{\delta[(p+K)^2-m^2]}{p^2-m^2}\{n_F(p+K,\mu) + n_F(p+K,-\mu)\} + \frac{\delta[p^2-m^2]}{(p+K)^2-m^2}\{n_F(p,\mu) + n_F(p,-\mu)\}\right] \quad (10)$$

The polarization tensor $\pi_{\mu\nu}^{\beta}(K,\mu)$ can generally be written in terms of the longitudinal and transverse components $\pi_L(k,\omega)$ and $\pi_T(k,\omega)$, respectively, such that it satisfies the relation

$$\pi_{\mu\nu}(K,\mu) = P_{\mu\nu}\pi_T(K,\mu)+ Q_{\mu\nu}\pi_L(K,\mu), \quad (11)$$

Whereas the polarization tensor corresponding to the transversely polarized wave is

$$P_{\mu\nu}=\tilde{g}_{\mu\nu} + \frac{\tilde{K}_\mu \tilde{K}_\nu}{k^2} \quad (12a)$$

and the polarization tensor corresponding to the longitudinally polarized wave (only possible in a statistical medium) is

$$Q_{\mu\nu} = -\frac{1}{K^2k^2}\left(k^2u_\mu + \omega\tilde{K}_\mu\right)\left(k^2u_\nu + \omega\tilde{K}_\nu\right) \quad (12b)$$

whereas,

$$\tilde{g}_{\mu\nu}=g_{\mu\nu} - u_\mu u_\nu \quad (12c)$$

and

$$\tilde{K}_\mu=K_\mu - \omega u_\mu \quad (12d)$$

Such that they satisfy the conditions:

$$P_\nu^\mu P_\alpha^\mu = P_\alpha^\mu \qquad Q_\alpha^\mu = Q_\nu^\mu Q_\alpha^\nu \qquad K_\mu P_\nu^\mu = 0 \qquad K_\mu Q_\nu^\mu = 0 \quad (13)$$

with

$$\Pi_L(\omega,k)=0 \quad (14)$$

in vacuum. In the absence of the longitudinal component of photon (in vacuum), Eq. (11) reduces to

$$\pi_{\mu\nu} = P_{\mu\nu} \qquad (15)$$

Showing that all the light is transversally polarized and $\varepsilon(K)=1$ and $\mu(K)=1$. Transversality of photon (Eq.(14) is the property associated with the masslessness of photon. When the photon acquires a plasma screening mass at nonzero temperature, it gives a nonzero contribution to the longitudinal component of the polarization tensor which shows the trapping of longitudinal component of electromagnetic waves in a medium. Nonzero parallel component in three dimensional space can still maintain the circular polarization in an isotropic medium and the possibility of trapping of light in a medium affects the transverse propagation in anisotropic medium in extreme statistical conditions. Due to the mass of the photon light slows down in the transverse direction by losing some energy to longitudinal direction.

Temperature and density corrections to QED parameters in a hot and dense medium are reviewed in the next section. The magnetic field is not included explicitly. However, in the presence of charged leptons at high temperatures, significantly large magnetic field is expected. In a closed system with isotropic matter distribution has a constant magnetic field at a constant temperature. Therefore, at a given temperature, magnetic field effect is incorporated through the potential energy contribution to the charged leptons and the energy of particles will be modified in the presence of the magnetic field as

$$E^2= p_0^{2=}p^2+m^2\pm\mu N+(2\ell+1)eB$$

Where ℓ correspond to the Landau level and B is the constant magnetic field. B can be replaced by the time varying magnetic field to incorporate the change in magnetic energy with time. We postpone the detailed study of the effect of different type of magnetic fields for now.

## 3. QED Parameters in a Medium

Thermal corrections to QED parameters can be written as a function of temperature T and chemical potential μ in the form of Masood's a,b,c, ….functions expressed as $a_i(m\beta,\mu)$ [5-22] and referred to as Masood's functions hereafter

$$\frac{\delta m}{m} \approx \frac{\alpha\pi T^2}{3m^2}\left[1-\frac{6}{\pi^2}c(m\beta,\mu)\right]+\frac{2\alpha}{\pi}\frac{T}{m}a(m\beta,\mu)-\frac{3\alpha}{\pi}b(m\beta,\mu). \qquad (16)$$

First term in bracket is a measure of thermal corrections due to the increase in kinetic energy due to the coupling of particles with radiation, whereas a, b and c functions are evolved from the integration of Fermi-Dirac distribution and vanishes at low temperature where the presence of hot fermions is negligible in the system.

$$a\left(m\beta,\pm\mu\right)=ln\left(1+e^{-\beta(m\pm\mu)}\right), \qquad (17a)$$

$$b\left(m\beta,\pm\mu\right)=\sum_{n=1}^{\infty}\left(-1\right)^{n}e^{\mp\beta\mu}Ei\left(-nm\beta\right), \qquad (17b)$$

$$c\left(m\beta,\pm\mu\right)=\sum_{n=1}^{\infty}\left(-1\right)^{n}\frac{e^{-n\beta(m\pm\mu)}}{n^{2}}. \qquad (17c)$$

Where, +(-)μ correspond to the chemical potential of fermion (antifermion) in the medium. The wavefunction renormalization constant of QED can be written as [5]

$$Z_2^{-1}(\beta)=Z_2^{-1}(T=0)-\frac{2\alpha}{\pi}\int_0^{\infty}\frac{dk}{k}n_B(k)-\frac{5\alpha}{\pi}b(\mathrm{m}\beta,\mu)$$
$$+\frac{\alpha T^2}{\pi v E^2}\ln\frac{1+v}{1-v}\left\{\frac{\pi^2}{6}+m\beta a(\mathrm{m}\beta,\mu)-c(\mathrm{m}\beta,\mu)\right\}, \qquad (18)$$

and the charge renormalization constant is calculated [5] as :

$$Z_3\cong 1+\frac{2e^2}{\pi^2}\left[\frac{ma(m\beta,\mu)}{\beta}-\frac{c(m\beta,\mu)}{\beta^2}+\frac{1}{4}\left(m^2+\frac{1}{3}\omega^2\right)b(m\beta,\mu)\right]. \qquad (19)$$

The photon in the medium develop a plasma screening mass which can be obtained from the longitudinal and transverse component of the vacuum polarization tensor $\Pi_L(0,k)$ and $\Pi_T(k,k)$ where $K^2=\omega^2-k^2$.

In this scheme of calculations, longitudinal and transverse components ($\Pi_L$ and $\Pi_T$, respectively) of vacuum polarization tensor $\Pi_{\mu\nu}$ play a crucial role in the calculation of the electromagnetic properties of a medium. The electromagnetic properties such as electric permittivity ε(K) and magnetic permeability μ(K). Other related properties of the medium including refractive index, propagation speed and the magnetic moment of different particles in the medium are studied by using these basic properties of the medium. Electric permittivity $\varepsilon(K)$ and the magnetic permeability $\mu(K)$ can be expressed [9] in terms of $\Pi_L$ and $\Pi_T$ such that:

$$\varepsilon(K)=1-\frac{\Pi_L}{K^2}, \qquad (20a)$$

$$\frac{1}{\mu(K)}=1+\frac{K^2\Pi_T-\omega^2\Pi_L}{k^2K^2}, \qquad (20b)$$

Such that

$$\Pi_L\cong\frac{4e^2}{\pi^2}\left(1-\frac{\omega^2}{\mathbf{k}^2}\right)\left[\left(1-\frac{\omega}{2\mathbf{k}}\ln\frac{\omega+\mathbf{k}}{\omega-\mathbf{k}}\right)\left(\frac{ma(m\beta,\mu)}{\beta}-\frac{c(m\beta,\mu)}{\beta^2}\right)\right.$$
$$\left.+\frac{1}{4}\left(2m^2-\omega^2+\frac{11\mathbf{k}^2+37\omega^2}{72}\right)b(m\beta,\mu)\right], \qquad (21a)$$

$$\Pi_T \cong \frac{2e^2}{\pi^2}\Big[\Big\{\frac{\omega^2}{\mathbf{k}^2}+\Big(1-\frac{\omega^2}{\mathbf{k}^2}\Big)\frac{\omega}{2\mathbf{k}}\ln\frac{\omega+\mathbf{k}}{\omega-\mathbf{k}}\Big\}\Big(\frac{ma(m\beta,\mu)}{\beta}-\frac{c(m\beta,\mu)}{\beta^2}\Big)$$
$$+\frac{1}{8}\Big(2m^2+\omega^2+\frac{107\omega^2+131\mathbf{k}^2}{72}\Big)b(m\beta,\mu)\Big]. \qquad (21b)$$

Whereas at extremely high temperatures, (at $\omega = k$ )

$$\lim_{k\to 0}\Pi_L(0,k) \equiv K^2{}_L \cong \frac{e^2T^2}{3} \qquad \text{(T>>m)} \qquad (22a)$$

$K_L$ correspond to Debye length. If we set $\omega = |k| = k$ with $k\to 0$, then,

$$\lim_{k\to 0}\Pi_T(k,\vec{k}) \equiv \omega^2{}_T \cong \frac{e^2T^2}{6} \qquad \text{(T>>m)} \qquad (22b)$$

Whereas in the relativistic plasma, it depends on temperature quadratically. Now the electric permittivity $\varepsilon(K)$ and magnetic permeability $\mu(K)$ of such a medium can be calculated from the longitudinal and transverse components [9] as:

$$\varepsilon(K) \cong 1-\frac{4e^2}{\pi^2K^2}\left(1-\frac{\omega^2}{\mathbf{k}^2}\right)\Big\{\left(1-\frac{\omega}{2\mathbf{k}}\ln\frac{\omega+\mathbf{k}}{\omega-\mathbf{k}}\right)\left(\frac{ma(m\beta,\mu)}{\beta}-\frac{c(m\beta,\mu)}{\beta^2}\right)$$
$$+\frac{1}{4}\left(2m^2-\omega^2+\frac{11k^2+37\omega^2}{72}\right)b(m\beta,\mu)\Big\}, \qquad (23a)$$

and

$$\frac{1}{\mu(K)} \cong 1-\frac{2e^2}{\pi^2k^2K^2}\Big[\omega^2\left\{\Big(1-\frac{\omega^2}{\mathbf{k}^4}-\left(1+\frac{\mathbf{k}^2}{\omega^2}\right)\left(1-\frac{\omega^2}{\mathbf{k}^2}\right)\frac{\omega}{2\mathbf{k}}\ln\frac{\omega+\mathbf{k}}{\omega-\mathbf{k}}\right\}$$
$$\times\left(\frac{ma(m\beta,\mu)}{\beta}-\frac{c(m\beta,\mu)}{\beta^2}\right)-\frac{1}{8}\left(6m^2-\omega^2+\frac{129\omega^2-109k^2}{72}\right)b(m\beta,\mu)\Big\}\Big]. \qquad (23b)$$

Eq. (21) and Eq. (23) show the dependence of the longitudinal and transverse components of the polarization as well as the electric permittivity and the magnetic permeability of the medium as a function of ω and k corresponding to the propagating electromagnetic waves.

Dependence of $\varepsilon(K)$ and $\mu(K)$ on temperature induces temperature dependence to the propagation velocity and the refractive index of the medium. These quantities also varies with the energy and the momentum of photon. This is a distinct feature of the extremely hot medium of the early universe that the refractive index will depend on the wave properties as well as the temperature of the medium. The speed of propagation of electromagnetic waves in such a medium can be expressed as

$$v_{prop} = \sqrt{\frac{1}{\varepsilon(K)\mu(K)}} . \qquad (24)$$

and the refractive index of the medium comes out to be

$$\text{n} = \frac{c}{v} = \sqrt{\frac{\varepsilon(K)\mu(K)}{\varepsilon_0(K)\mu_0(K)}} \qquad (25)$$

and the Debye shielding length comes out to be

$$\lambda_D \cong \frac{e^2 T^2}{3}$$

In the early universe for an extremely large momentum of photon. It is known that the Debye length in classical plasma depends on temperature as

$$\lambda_D = \left(\frac{\varepsilon\, k_B T}{\sum_{j=1}^{N} n_j^0\, q_j^2}\right)^{1/2}$$

Showing the difference between classical and relativistic plasma.

## 4. Propagation of light in the Early Universe

It is well-known from thermal history of the universe that temperatures of the universe were extremely higher than the chemical potential and the valid limit of temperature is T >> µ. In this section, we evaluate all of Masood's functions for extremely high temperatures. Dominant thermal contribution comes from the interaction with the radiation at the equilibrium temperature T and fermions at the same temperature. The values of $a_i(m\beta, \mu)$ for extremely high temperature, only $c(m\beta, \mu)$ term contributes, giving

$$c(m\beta, \mu) = -\frac{\pi^2}{12}$$

And in the extremely high temperature limit (T>>m>>µ) with ignorable density the longitudinal and transverse components of the vacuum polarization tensor can be written as

$$\varepsilon(K) \cong 1 - \left(1 - \frac{\omega^2}{\mathbf{k}^2}\right)\left(1 - \frac{\omega}{2\mathbf{k}} \ln \frac{\omega + \mathbf{k}}{\omega - \mathbf{k}}\right)\frac{4\pi\alpha T^2}{3K^2}, \qquad (26a)$$

$$\frac{1}{\mu(K)} \cong 1 + \frac{2\pi\alpha T^2}{3K^2}\frac{\omega^2}{k^2}\left\{\left(1 - \frac{\omega^4}{\mathbf{k}^4} - \left(1 + \frac{\mathbf{k}^2}{\omega^2}\right)\left(1 - \frac{\omega^2}{\mathbf{k}^2}\right)\frac{\omega}{2\mathbf{k}} \ln \frac{\omega + \mathbf{k}}{\omega - \mathbf{k}}\right\} \qquad (26b)$$

For large values of k, for the relativistic systems, we can write $\varepsilon(K)$ and $\frac{1}{\mu(K)}$ as

$$\varepsilon(K) = 1 - \frac{e^2T^2}{3K^2}, \qquad (27a)$$

$$\frac{1}{\mu(K)} = 1 + \frac{(K^2 - 2\omega^2)e^2T^2}{6k^2K^2}, \qquad (27b)$$

Eqns. (26) and (27) can also be evaluated for two extreme conditions based on the properties of light as

(i ) For large energy of photons, $\omega \gg k$

$$\pi_L = -\frac{\omega^2e^2T^2}{3k^2} \qquad \pi_T = \frac{\omega^2e^2T^2}{6k^2} \qquad (28)$$

$$\varepsilon(k) = 1 + \frac{\omega^2e^2T^2}{3k^4} \qquad \frac{1}{\mu(k)} \cong \frac{\omega^4e^2T^2}{3k^4K^2} \qquad (29)$$

$$v_{prop} = \sqrt{\frac{\omega^4e^2T^2}{K^2(3k^4 + \omega^2e^2T^2)}}\,. \qquad (30)$$

and the refractive index of the medium comes out to be

$$\mathrm{n} = \frac{c}{v} = \sqrt{\frac{K^2(3k^4 + \omega^2e^2T^2)}{\omega^4e^2T^2}} \qquad (31)$$

(ii) For large momentum of photons, $\omega \ll k$

$$\pi_L = \frac{e^2T^2}{3} \qquad \pi_T = \frac{e^2T^2}{6} \qquad (32)$$

$$\varepsilon(k) = 1 - \frac{e^2T^2}{6k^2} \qquad \frac{1}{\mu(k)} \cong \frac{e^2T^2}{6K^2} \qquad (33)$$

Thermal correction to the speed of the propagation of light in such a medium is

$$v_{prop} = \sqrt{\frac{e^2T^2}{(6k^2 - e^2T^2)}}\,. \qquad (34a)$$

And the refractive index of this medium is given by

$$n = \sqrt{\frac{(6k^2 - e^2T^2)}{e^2T^2}}\,. \qquad (34b)$$

Eqn, (34) puts some natural limit on the values of k. $k^2=\frac{e^2T^2}{6k^2}$ is not physically allowed as it will give infinite speed and zero refractive index. These equations cannot be true for T=0 as well. The values of electric permittivity and magnetic permeability depend on the values of the plasmon frequency and the wave number at a given temperature and $v_{prop}$ and the refractive index n also become a function of temperature and the energy and momentum of photon.. It is also to be noticed that the thermal contributions (in Eqs. (25)) start at T > 0.5 MeV. So the electromagnetic properties of the medium cannot see any thermal effect after the nucleosynthesis [13, 14] in the early universe. This limit below the neutrino decoupling temperature indicating that the nucleosynthesis started right after the temperature of the universe dropped below the neutrino decoupling temperature and the neutrino capture contributed to beta processes in the production of electrons. All the calculations in this paper are relevant for the temperatures below the neutrino decoupling temperatures and higher than the electron mass.

Existence of hot electrons in a medium at such temperatures insures a significant effect on physically measureable values of electron mass, charge and concentration of electrons in a medium which leads to the change in the electromagnetic properties in terms of magnetic moments of leptons, electric permittivity and magnetic permeability of the medium. The electromagnetic properties of a medium then effect differently on the particles that propagate through this medium.

## 5. Magnetic Moment of charged particles

Relativistically moving charged particles have an associated electric field and their relativistic motion at high temperatures create weak non-negligible currents in extreme situations in a medium. When charged particles are accelerated in a medium and a continuous change in energy occurs through acceleration of particles along with the change in temperature of the system, an associated magnetic field is generated. In such an electromagnetic system, there are localized electromagnetic fields that are associated with the distribution of charged particles in the medium. Magnetic moment is associated with charge and mass of leptons. Lighter particles have large magnetic moment effect and thermal contribution is higher for the lighter particles as temperature is always compared to the mass of the particles for thermal effects. Renormalization scheme gives a change in mass in Eq. (2). The charge of electron is not affected until the temperature is extremely high as indicated in Eq. (5).

Magnetic moment is simply calculated from the change in mass in thermal background given as

$$a\mu = \left(\frac{\alpha}{2\pi} - \frac{2}{3}\frac{\delta m}{m}\right)\mu B \qquad (35)$$

$$\frac{\delta m}{m} = \frac{\delta m}{m}(vacuum) + \frac{\delta m}{m}(T, \mu) \qquad (36)$$

$\mu_B$ is the unit of magnetic moment called Bohr Magneton. So the statistical corrections to the magnetic moment of electron is directly proportional to the statistical selfmass corrections, given in Eq. (16).

**Generalization of results**

A straightforward generalization of all the above results can be done by evaluating Masood's functions $a_i$ such that $a_1 = a(m\beta, \mu), \; a_2 = b(m\beta, \mu), \; a_3 = c(m\beta, \mu),$ and so on.
All the Masood's statistical functions $a_i$ correspond to electron background contributions in the above equations and always correspond to fermion background contribution for T > µ. Evaluating the above equations, we just consider the electron background as m is taken as the electron mass. Thus these functions correspond to the electron background for temperatures higher than the electron mass. Positrons contribution in the same medium can be expressed by replacing µ with -µ. Everything else remains unchanged. Net contributions from the CP symmetric background can be obtained by taking an average of particle and antiparticle background contributions and a net background contributions can be obtained by replacing $a_i$ functions by the corresponding difference functions, giving [5]

$$a_{avg}(m\beta, \mu) = \frac{1}{2}[a(m\beta, \mu) + a(m\beta, -\mu)] \qquad (37a)$$

$$a_{net}(m\beta, \mu) = \frac{1}{2}[a(m\beta, \mu) - a(m\beta, -\mu)] \qquad (37b)$$

such that , in the limit $T >> \mu$, chemical potential contribution is ignorable and hence the thermal effects dominate, such that

$$a_{net}(m_\ell \beta, \mu) \to 0 \qquad (38a)$$

$$a_{avg} m_\ell \beta, \mu) \to a_i(m_\ell \beta,) \qquad (38b)$$

So the average background contribution can be obtained by the corresponding functions to Eqns. (17)

$$a_{avg}(m_\ell \beta, \mu) \cong \frac{1}{2} \ell n\left[\left(1 + e^{-\beta(m_\ell - \mu)}\right)\left(1 + e^{-\beta(m_\ell + \mu)}\right)\right] \qquad (39a)$$

$$b_{avg}(m_\ell \beta, \mu) \cong \sum_{n=1}^{\infty} (-1)^n \cosh(n\beta\mu)\, \mathrm{Ei}(-nm_\ell \beta) \qquad (39b)$$

$$c_{avg}(m_\ell \beta, \mu) \cong \sum_{n=1}^{\infty} (-1)^n \cosh(n\beta\mu) \frac{e^{-n\beta m_\ell}}{n^2} \qquad (39c)$$

It can be easily seen that at extremely high temperatures, the fermion contribution from the medium is controlled by $c(m\beta, \mu)$ as it can be easily shown that $c(m\beta, \mu) = -\frac{\pi^2}{12}$ .

## 6. Results and Discussions

It can be seen from Eqs. (16-19) that the background contribution to electron mass, wavefunction and charge comes directly from the hot background as a function of temperature which brings in $(T/m)^2$, as the dominant contribution. The photon background contribution is always proportional to $(T/m)^2$. Thermal contribution to electron mass, wavefunction and charge of electron at high temperature is plotted in figure 2 showing a comparison of thermal contribution to all the three renormalization parameters of QED. They have a similar dependence on temperature with different coefficients. Thermal contribution to the electron selfmass is a couple of orders of magnitude greater than the electron charge and the coupling constant of QED. It can be seen from Eqns. (16-19) that the wavefunction correction is much smaller than the electron mass or even the charge of electron. It can be easily seen that the mass contribution dominates over the charge contribution because electron mass has radiative corrections by its interaction with the radiation (photon) in the background. However, charge does not see the photon and thermal contribution is only due to the fermion background. Thermal corrections to the electron wavefunction are very small even at T > m. However, the charge of electron is not affected by the background at low temperatures (T<m) because massless photons do not interact among themselves. At T>m photon acquires dynamically generated mass which increases with temperature in the presence of high concentration of electrons in a medium. The QED parameters indicating properties of electrons at high temperature are plotted in Figure (2).

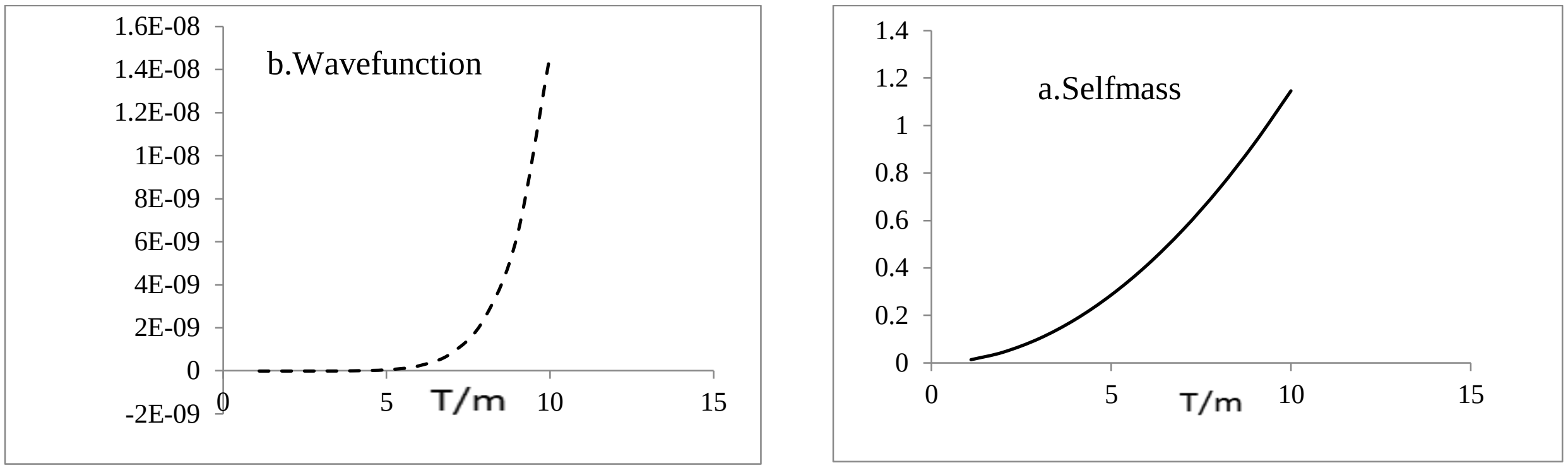

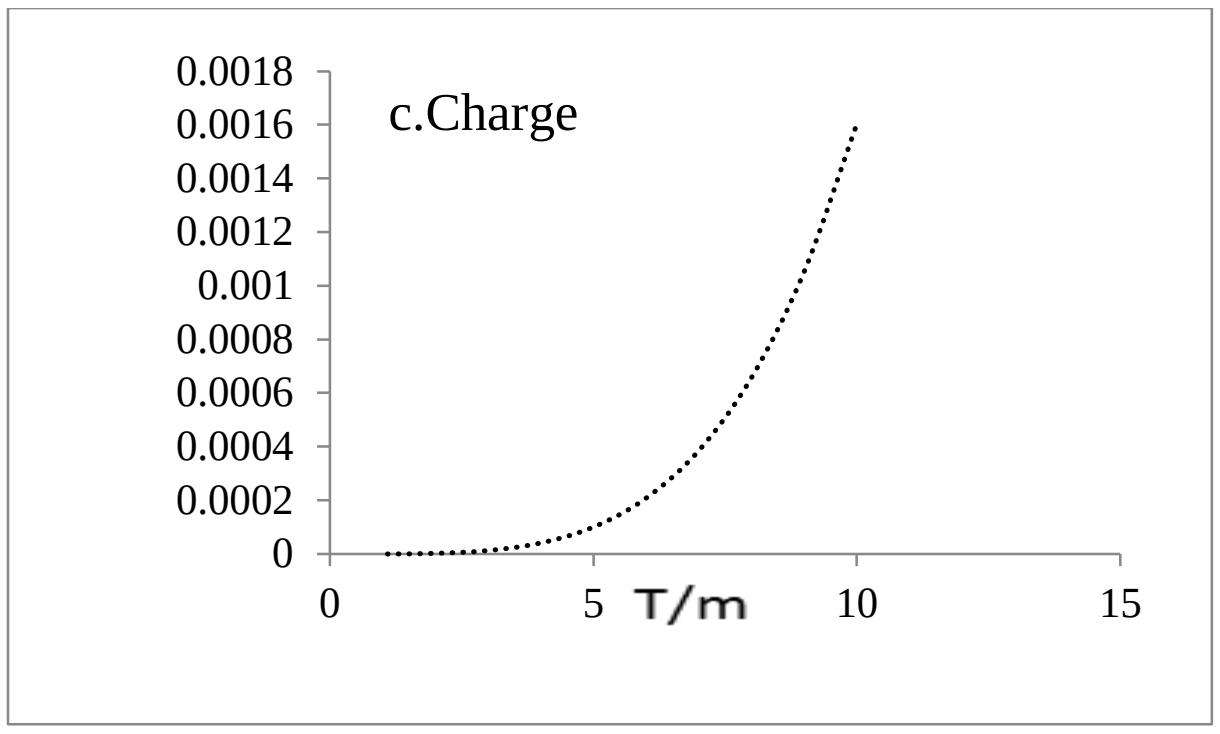

*Figure2: Properties of electrons as a function of temperature (in units of m). Thermal corrections to a).electron mass; b).wavefunction; and c).charge.*

Selfmass corrections to electron affect the magnetic moment of electron which is also proportional to the $T^2/m^2$. Figure (3) gives the magnetic moment of electron as a function of temperature as the electron mass keeps on increasing with temperature. However, the presence of charged fermions in the background affects the electromagnetic properties of electron also. Figure 3 plots the thermal contributions to the magnetic moment as compared to the thermal contribution to the mass of electron.

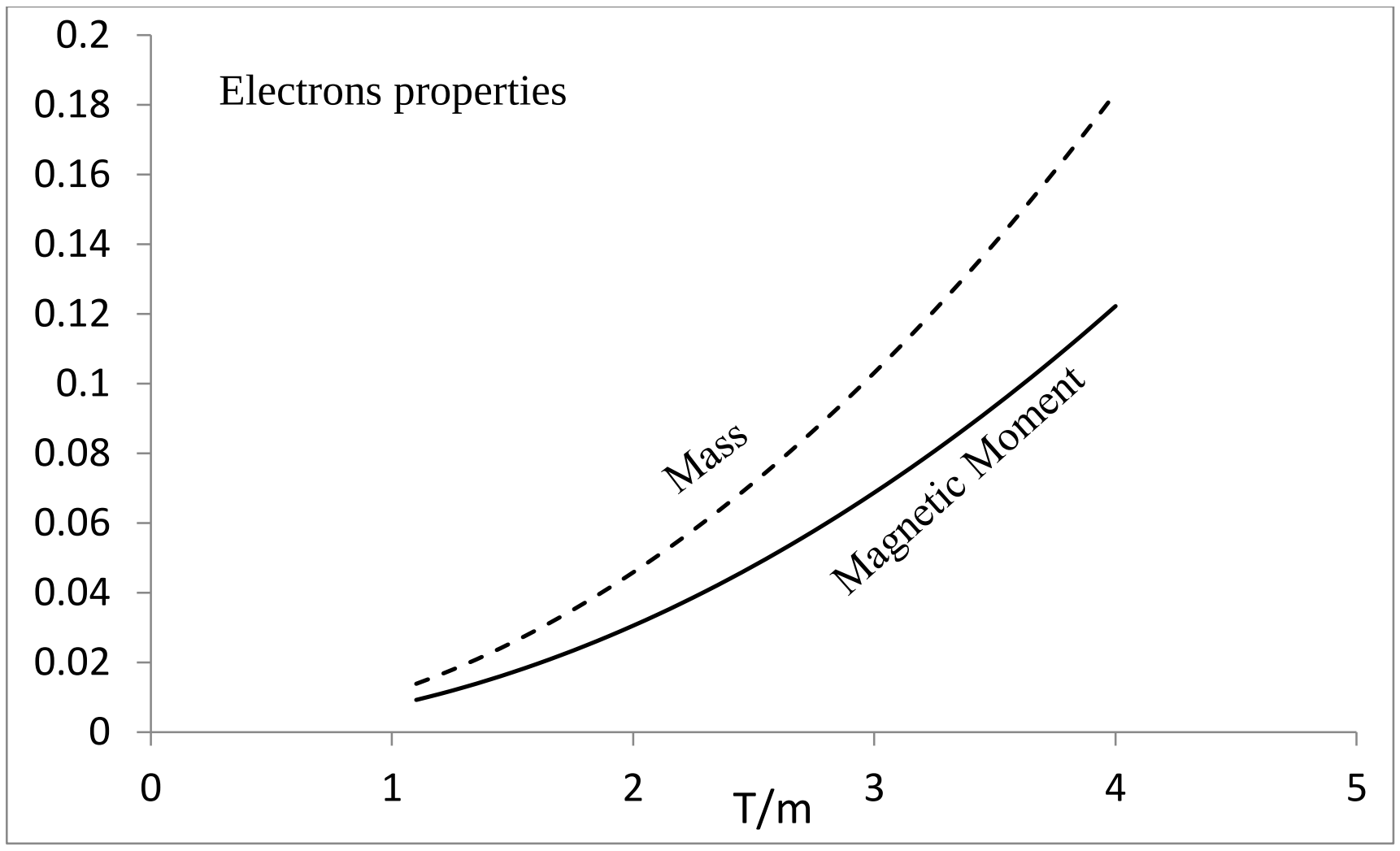

*Figure 3: A comparison of temperature dependent selfmass and the corresponding magnetic moment of electron shows that the behavior of magnetic dipole moment at finite temperature s exactly similar to the mass of electron at high temperatures, i.e; T>m*

Magnetic moment is a form factor and is a property of mass and charge whereas neutrino is a massless neutral particle. So the nonzero magnetic moment can only be obtained through the weak interaction of neutrino with the corresponding charged lepton as shown in Figure (1). In the standard model massless and neutral neutrino cannot interact with the magnetic field and exhibits zero magnetic moment. It is a higher order process if neutrino has some mass. In this way, the magnetic moment of neutrino will depend on the extension of the standard model and will be a model dependent quantity. We just consider the minimally extended standard model with the neutrino mass as 1MeV just to compare thermal mass of electron and the magnetic dipole moment of neutrino as a function of temperature. This comparison is done at extremely high temperatures. Induced magnetic moment of neutrinos is plotted in Figure 4 where the magnetic moment of neutrino is plotted as a function of temperature for the upper limit of the mass of electron type neutrino around 1 eV. In this first order correction, the minimal standard model obeying the conservation of the individual lepton number is considered.

Temperature correction is suppressed for neutrino because the temperatures is compared to the mass of W, instead of electron as W boson is the loop partner of electron in the bubble diagram. Therefore, the background contribution to the magnetic moment of neutrino is induced as almost $10^{-10}$ times the corresponding contributions to electron mass which is exactly of the order of ($m^2/M^2$) and is of the order of the square of the ratios between the mass of electron and the mass of W boson. Figure 4 shows that this induced magnetic moment of neutrino in the minimal standard model is negligibly small being the higher order effect.

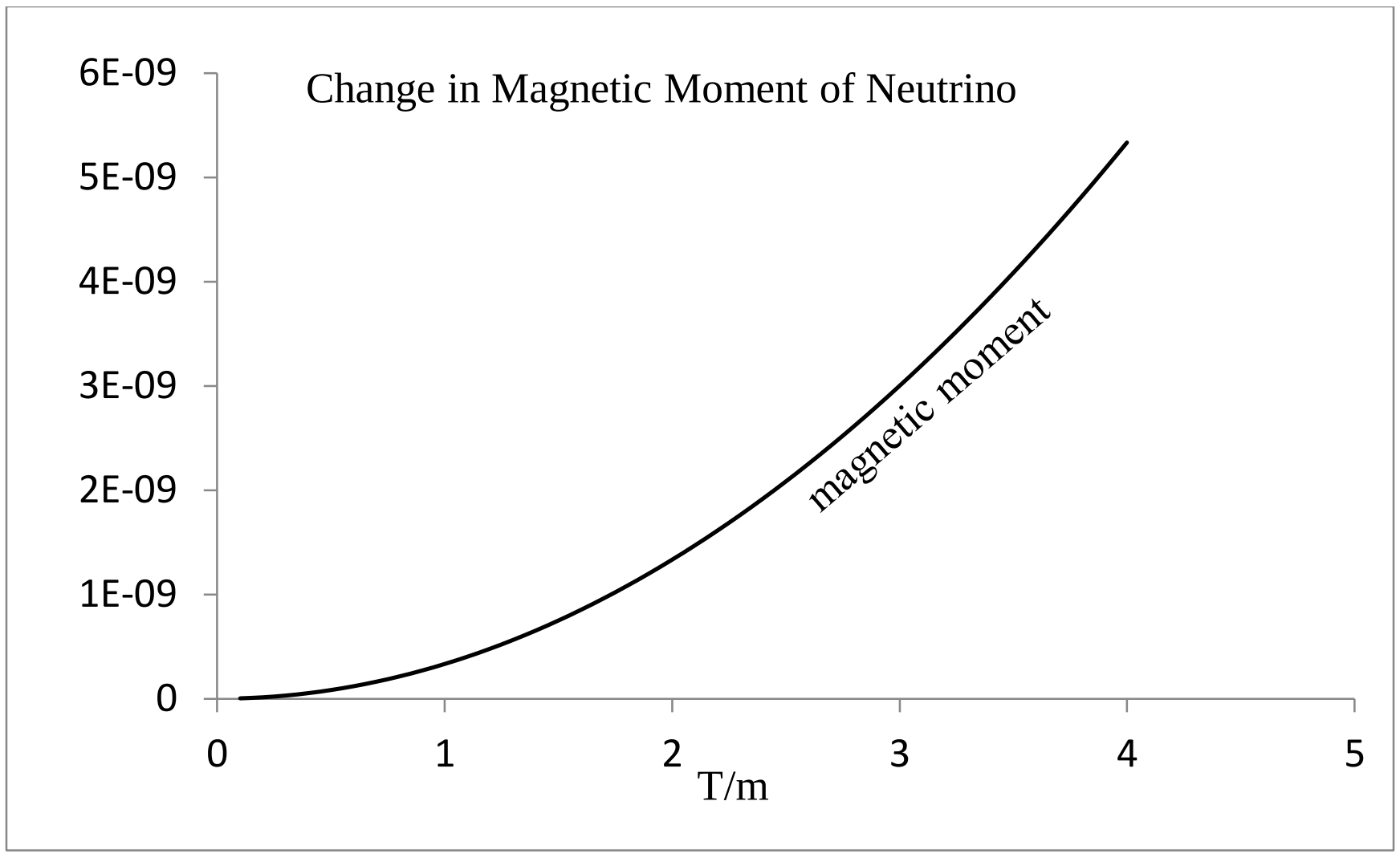

*Figure 4: The ratio of (thermal) electrons background corrections to the magnetic moment of electronic type neutrino with the corresponding vacuum value of the magnetic moment is plotted as a function of temperature below the neutrino decoupling temperature to extract the pure background effect.*

An explicit comparison of all these values is given in Table 1. It shows the values of all of the QED parameters such as electron properties and the magnetic moment of electron for a given value of temperature including the induced magnetic moment of neutrino. However, it is clear from the last column of the table that the ratio of the magnetic moment of electron type neutrino in lepton number conserving minimally extended standard model with that of electron is constant which is actually proportional to the ratio between their masses.

| ***T/m*** | ***δm/m*** | ***Δe/e*** | ***Electron Dipole moment*** | ***Neutrino Dipole moment*** | ***Ratio of Magnetic Moments*** |
|---|---|---|---|---|---|
| *1.1* | *0.013879* | *2.34E-07* | *0.009252* | *4.03333E-10* | *4.36E-08* |
| *1.5* | *0.025808* | *8.1E-07* | *0.017205* | *7.5E-10* | *4.36E-08* |
| *2* | *0.04588* | *2.56E-06* | *0.030587* | *1.33333E-09* | *4.36E-08* |
| *2.5* | *0.071688* | *6.25E-06* | *0.047792* | *2.08333E-09* | *4.36E-08* |
| *3* | *0.10323* | *1.3E-05* | *0.06882* | *0.000000003* | *4.36E-08* |

*Table 1: Electron selfmass, charge, magnetic moment and neutrino magnetic moment have been evaluated for the same values of temperatures in the units of electron mass and below the coupling temperatures. Chemical potential is ignored at this point to keep this comparison simple.*

Magnetic moment is a property of mass. Thermal corrections to the magnetic dipole moment of neutrino are actually due to its swallowed mass in thermal background. This expected behavior is demonstrated in a plot of electron mass and its corresponding modification in the dipole moment as a function of temperature. Figure 3 shows a clear demonstration of this behavior of electron. It is also interesting to note the vacuum polarization contribution due to the presence of fermions in the background for $T > m$. Neutrino magnetic moment can be induced the same way even if the neutrinos are not decoupled (below 2 MeV) as they acquire the induced magnetic dipole moment by electrons which can occur even if neutrino concentration is lower but there are enough electrons in the medium..

**Properties of photons at high temperature**

Photon, as a quanta of energy acquires nonzero mass in a medium with an abundance of electrons with extremely high energy at high temperature. Electromagnetic interaction of photon with electrons in a medium gives a dynamically generated mass to photon which can be treated as the screening mass and the Debye shielding turns the medium into an electron-photon relativistic plasma under suitable conditions. Since the photon is massless at lower temperature and density and the phase transition in a medium occurs at temperatures greater than 1 MeV where the fermion background starts to contribute to the self-energy of photon which leads to Debye shielding due to the dynamically generated mass of photons. The behavior of dynamically generated screening mass and Debye shielding length as a function of temperature is given in Figure 5. Debye screening length decreases with the increase in temperature and with the increase in screening mass of photon. At the temperature around 3-4 Mev, the Debye length decreases, compared to the screening mass itself. Debye length correspond to the potential energy and decrease in the potential energy is expected with temperature.

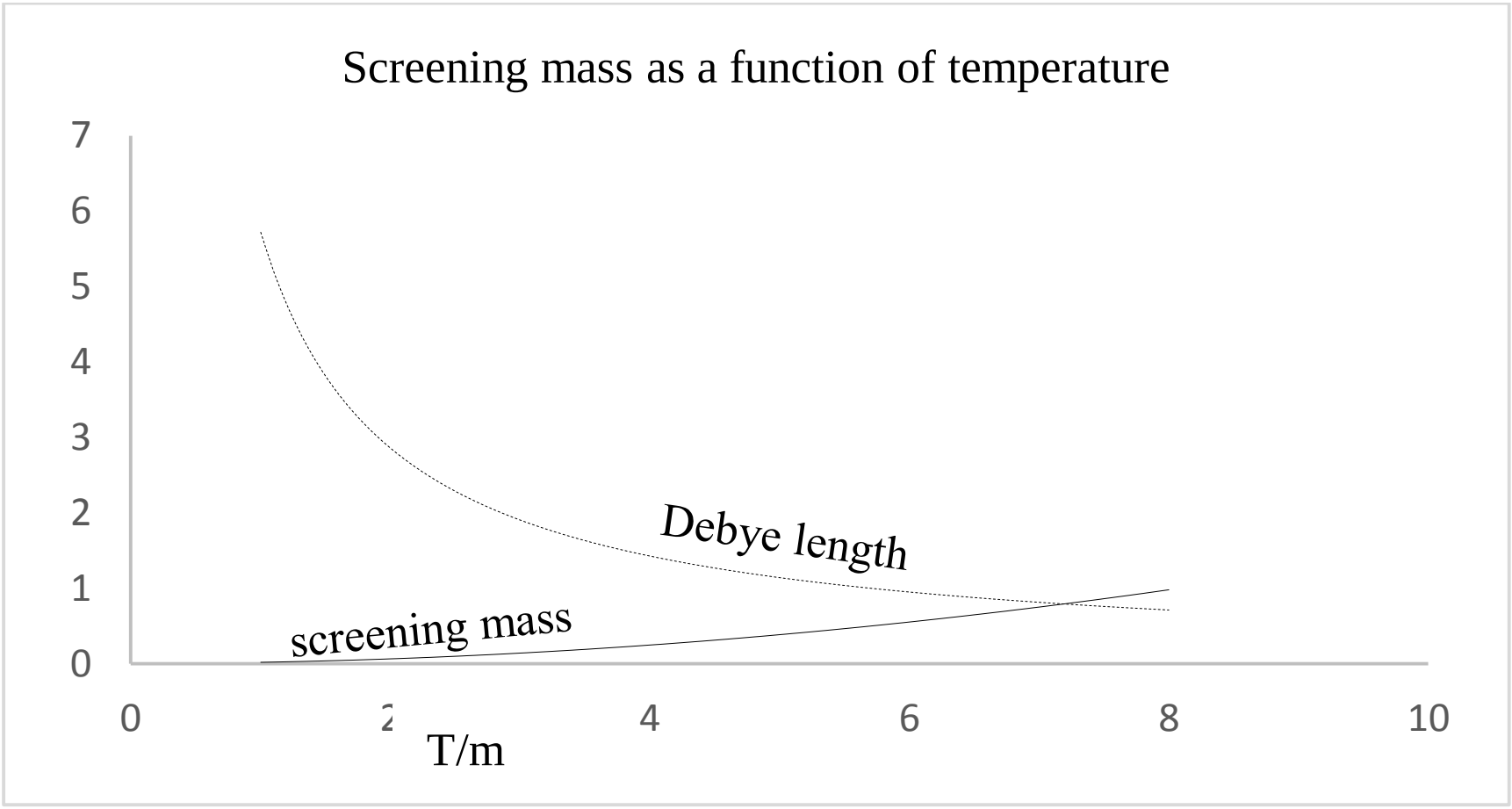

.

*Figure 5: Plot of plasma screening mass (solid line) and the corresponding Debye shielding length (broken line) as a function of temperature in units of electron mass.*

Number of particles in the Debye sphere can be calculated from Debye shielding length $\lambda_D$ , if the number density remains unchanged.

$$N_D = 4/3\ \pi\ n_0\ \lambda^3_D$$

Thus the decrease in $\lambda_D$ with temperature is associated with the number density of the universe. The composition of the universe changes with temperature and so does the $N_D$ due to the change in mass as well as the other parameters of the theory such as $\lambda_D$ and the propagation velocities.

We consider the small longitudinal component and the magnitude of the wave-vector k is much larger than the frequency ω making it a very high energy wave.

However the velocity of waves in a medium grows quickly with temperature whereas the refractive index decreases relatively slowly with temperature in such a medium.

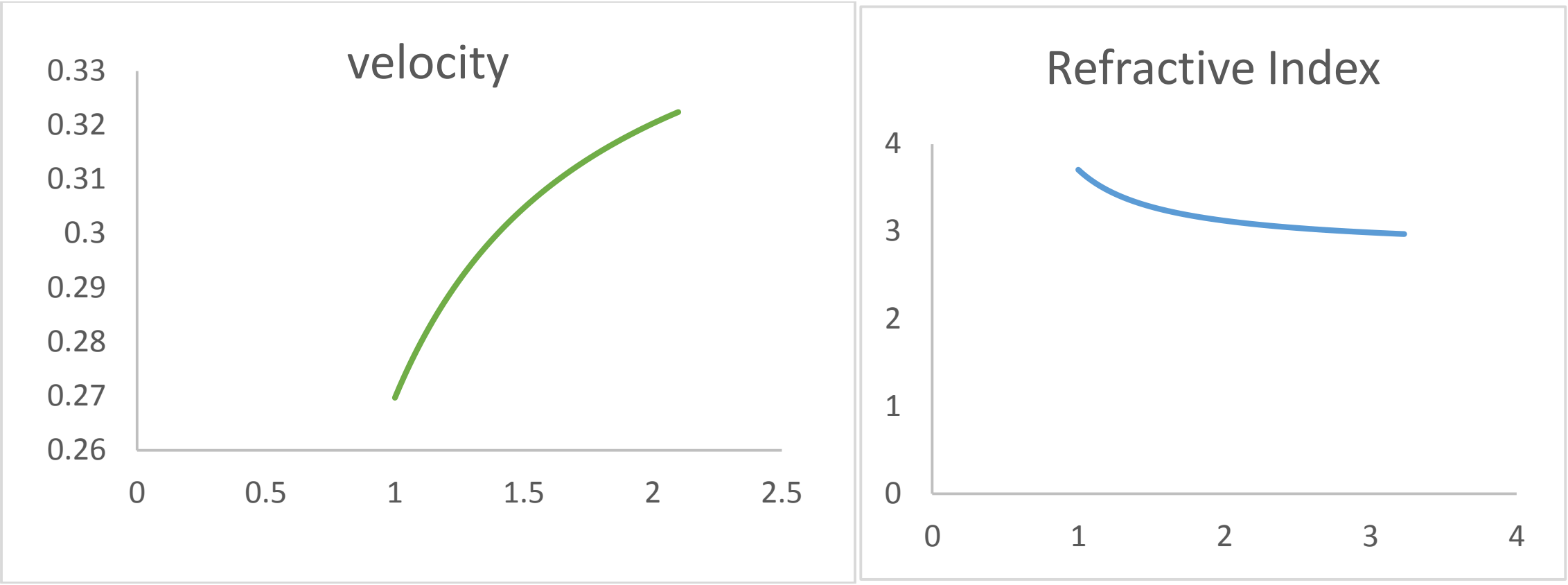

*Figure 6: Plot of propagation velocity and the refractive index in an extremely hot medium.*

The above results indicate that QED can be used as the only theory below the neutrino decoupling temperature and that is 2 MeV. As soon as the temperature of the system crosses this limit, concentration of neutrinos become significant and the weak interactions come in to play, some of the energy is used in weak interaction and is not ignorable with the rise of temperature. It can be clearly seen that the thermal corrections lead to quadratic increase in the measureable values of physical parameters of the theory [Figs. 2-4]. However the interaction based bulk properties such as the propagation speed, magnetic moment and the screening mass has different behavior. This difference in behavior balances some of the effects and keep the possibility of existence of physical systems at extreme temperatures (kinetic energies) and large densities or chemical potential (potential energies) and magnetic field help to develop an equilibrium and maintain the system.

Properties of neutrinos are related to the massive neutrinos which opens up a whole list of possible extensions of the standard model to accommodate massive neutrinos. Thermal effects on the properties of neutrinos are highly model dependent [20-23] and can only be calculated individually for every model. Even the Dirac and Majorana mass will contribute differently to the magnetic moment.
It can be clearly seen that all the above equations of section 4 reproduce the already existing results in the tranversality limit where $\omega^2 \approx k^2$, we get propagation velocity, electric permittivity and magnetic permeability are all equal to unity.

## REFERENCES AND FOOTNOTES